\documentclass[%
 aip,
rsi,%
 amsmath,amssymb,
 reprint,%
]{revtex4-1}

\usepackage{graphicx}
\usepackage{dcolumn}
\usepackage{bm}
\usepackage{amsmath}
\usepackage{siunitx}
\usepackage{braket}
\usepackage{upgreek}
\begin{document}

\preprint{AIP/123-QED}

\title{High fidelity field stop collection for polarization-entangled photon pair sources}

\author{Alexander Lohrmann}
\affiliation{%
Centre for Quantum Technologies, National University of Singapore, 3 Science Drive 2, S117543, Singapore\\
}%

\author{Aitor Villar}
\affiliation{%
Centre for Quantum Technologies, National University of Singapore, 3 Science Drive 2, S117543, Singapore\\
}%

\author{Arian Stolk}
\affiliation{%
Centre for Quantum Technologies, National University of Singapore, 3 Science Drive 2, S117543, Singapore\\
}%
\affiliation{%
Currently with QuTech, Delft University of Technology, PO Box 5046, 2600 GA Delft, The Netherlands\\
}%

\author{Alexander Ling}
\affiliation{%
Centre for Quantum Technologies, National University of Singapore, 3 Science Drive 2, S117543, Singapore\\
}%
\affiliation{%
Physics Department, National University of Singapore, 2 Science Drive 3, S117542, Singapore
}%

\date{\today}

\begin{abstract}We present an experimental demonstration of a bright and high fidelity polarization entangled photon pair source. 
The source is constructed using two critically phase matched $\beta$-Barium Borate crystals with parallel optical axes and photon pairs are collected after filtering with a circular field stop.
Near-unity fidelities are obtained with detected pair rates exceeding \SI{100000}{pairs/\s/\mW}, approaching the brightness of practical quasi-phase matched entangled photon sources. We find that the brightness scales linearly with the crystal length. We present models supporting the experimental data and propose strategies for further improvement.
The source design is a promising candidate for emerging quantum applications outside of laboratory environments.
\end{abstract}

\keywords{Entangled photon sources, spatial phase, critical phase matching}
\maketitle

\begin{figure*}[htb]
\centering
\includegraphics[scale = 0.8]{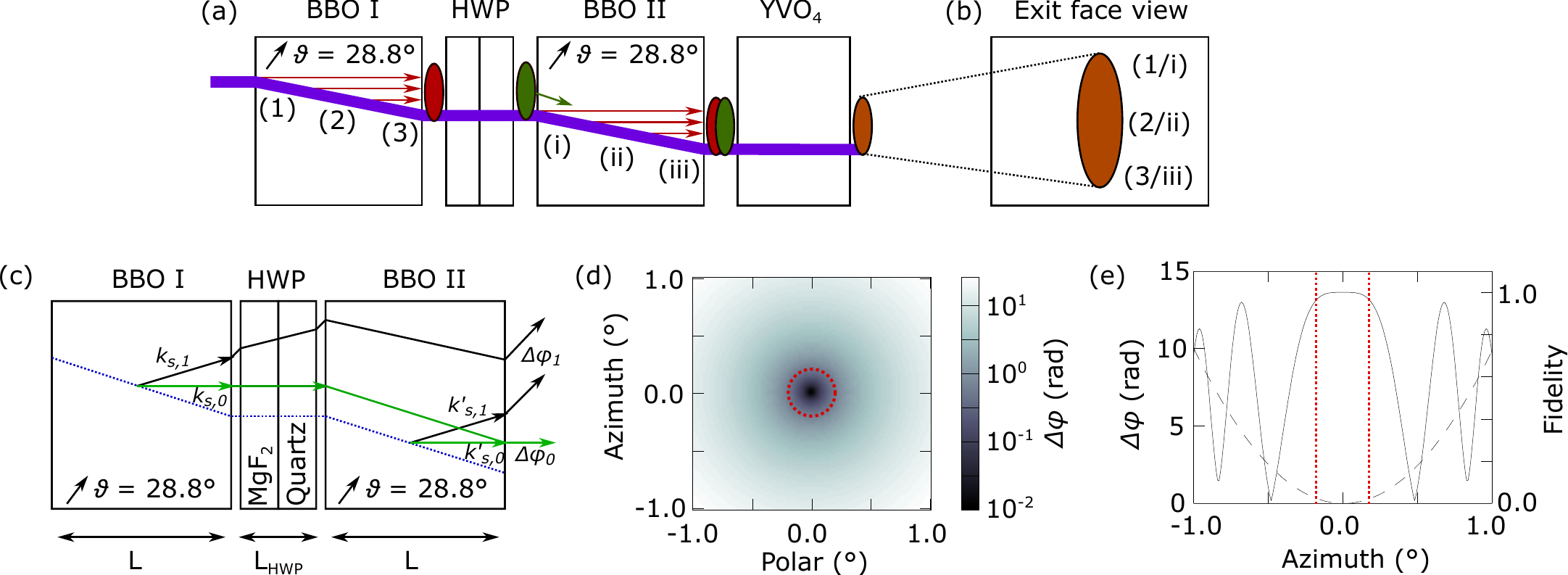}
\caption{(a) Basic principle of the parallel crystal configuration. The numbers 1-3 and i-iii indicate three possible SPDC generation positions for each crystal. The colors indicate the different photon polarizations (green: $\ket{VV}$, red: $\ket{HH}$). (b) Sketch of the SPDC spatial distribution at the exit face of the second crystal. The numbers 1-3 and i-iii indicate the photon origin as depicted in (a). (c) Signal (idler omitted for clarity) ray propagation diagram for collinear (green) and non-collinear emission (black). The rays in the figure indicate the light propagation (Poynting vector) for the initial $k$-vectors ($k_{s,1}$ and $k_{s,0}$ from the first crystal and $k^\prime_{s,1}$ and $k^\prime_{s,0}$ from the second crystal).”
 Pump indicated in blue. The angle dependence of the phase difference originates from the effective path length difference. (d) $\Delta\varphi$ as a function of the polar opening angles for the signal ray in air for every point in (b) for a source based on 6~mm BBO crystals. The area inside the red circle indicates a region of approximately constant fidelity. (e) Phase difference (dashed line) and fidelity of $\ket{\Phi^-}$ (solid line) as a function of the vertical polar emission angle (upwards in (a)) extracted from (d) when the horizontal polar angle is 0$^\circ$. The state transitions between $\ket{\Phi^-}$ ($F=1$) and $\ket{\Phi^+} ($F=0$)$. The red dotted lines indicate a region of approximately constant phase equivalent to (d). (Color online)}
\label{fig:schematic}
\end{figure*}

Entangled photon pairs lie at the heart of many emerging quantum technologies, such as quantum communication, key distribution and teleportation \cite{gisin2007quantum,horodecki2009quantum}. An on-going research problem is to develop sources of entangled photons that are bright and, at the same time, robust enough to be suitable for long term operation outside of laboratory environments. The main method of generating entangled photons is based on spontaneous parametric downconversion (SPDC) in nonlinear crystals.

One of the important considerations in the design of entangled photon sources is the effective collection angle of SPDC photons. This is critical because the angle dependent phase can degrade the desired entangled state. To negate this effect, many source designs\cite{trojek2008collinear,steinlechner2012high,steinlechner2013phase} and applications\cite{yin2017satellite,steinlechner2017distribution} use single-mode collection or spatial filtering. This, however, strongly reduces the source brightness. Collection through a field stop, on the other hand, may allow for high brightness and fidelity, if the angle dependent phase is taken into consideration
\cite{kwiat1999ultrabright,kim2006phase, altepeter2005phase}. Previous work in this direction has been restricted to the use of thin crystals\cite{ altepeter2005phase,kim2006phase,hegazy2015tunable, hegazy2017relative}. In this paper, we report on the collection of high fidelity entangled photons through a field stop when using thick crystals.



This result utilizes a source based on type-I, collinear, non-degenerate critical phase matching using two $\beta$-Barium Borate (BBO) crystals with parallel optical axes\cite{villar2018experimental}. 
The parallel axes approach leads to an almost perfect spatial overlap of the emission modes enabling the detection of maximally entangled photon pairs without single-mode filtering. We demonstrate pair rates of up to \SI{100000}{pairs/s/mW} with near-unity fidelity. Furthermore, by relaxing the collection conditions, we achieve rates of  \SI{400000}{pairs/s/mW} with QKD compatible fidelity.
 
\begin{figure}[b]
\centering
\includegraphics[scale = 1]{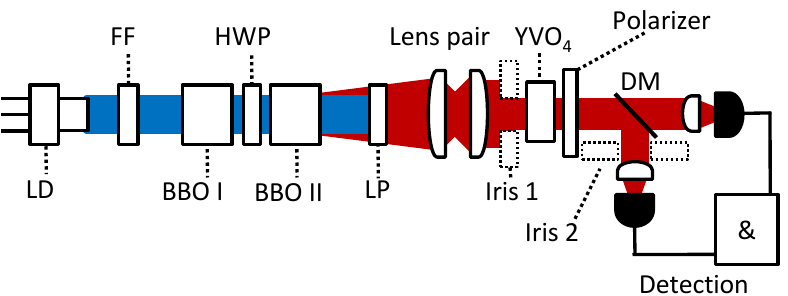}
\caption{Sketch of the experimental setup. LD: laser diode, FF: fluorescence filter, HWP: half-wave plate, LP: long-pass filter, DM: dichroic mirror. An iris is used to control the opening angle of the SPDC light that reaches the detectors. It can be placed either before splitting signal and idler (Iris 1) or after the dichroic mirror in the signal arm (Iris 2).  The YVO$_4$ (length: 3.6~mm) is placed in the collimated beam to avoid an additional angle dependence of $\Delta\varphi$. (Color online)}
\label{fig:setup}
\end{figure}

The source design is sketched in Fig.~\ref{fig:schematic}(a). The pump with a vertical (extraordinary) polarization undergoes a walk-off within the first BBO crystal. Due to the type-I phase matching, photon pairs with horizontal (ordinary) polarization ($\ket{H_sH_i}_1$) are generated. An achromatic half-wave plate rotates the polarization of these photon pairs to the orthogonal state ($\ket{H_sH_i}_1 \rightarrow \ket{V_sV_i}_1$) while leaving the pump polarization unaffected. 

In the second crystal the pump again generates $\ket{H_sH_i}_2$ photons. The photon pair from the first crystal has extraordinary polarization in the second crystal and therefore undergoes a walk-off in the same direction as the pump. This is the reason for the almost perfect spatial overlap of the SPDC emission modes from the two crystals. A residual spatial mismatch between the two modes exists, due to the different walk-off angles of pump and SPDC photons\cite{villar2018experimental} (walk-off angle difference: $\Delta \rho \leq 6\%$), but this is negligible and only adds a spatially constant phase difference between the horizontal and vertical pairs. 


\begin{figure}[tb]
\centering
\includegraphics[scale = 1]{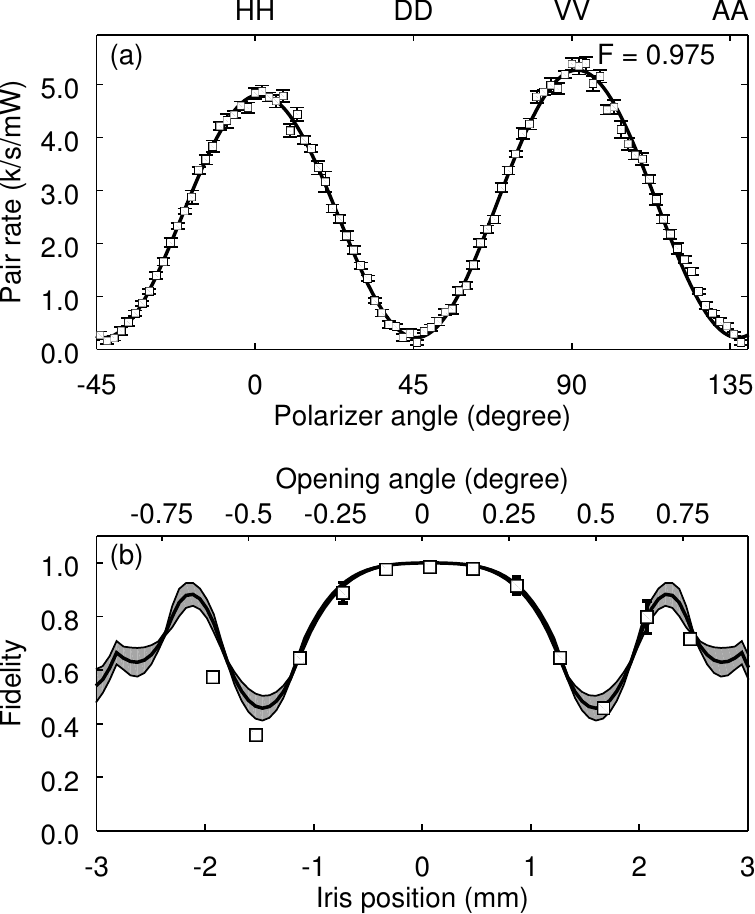}
\caption{ (a) Detected photon pair rate when using a single polarizer acting simultaneously on signal and idler photons. In this case the iris was centered in the signal arm. The diameter of the iris was $0.65\pm0.10$~mm. The fidelity was $97.5\pm0.3$\%. (b) Fidelity with respect to vertical iris translation. The solid line shows Eq.~\ref{eq:phi} (see supplementary material), where $\vec{\alpha}$ is assumed to be linear in iris position. The gray shaded area indicates the uncertainty of 1$\sigma$ due to the finite iris diameter. Squares are experimental data.}
\label{fig:azimuth}
\end{figure}

To understand how the spatial overlap enables high fidelity field stop collection, spatial considerations need to be added to the well-known phase compensation scheme\cite{trojek2008collinear}. In general, the state generated after the two BBO crystals can be written as, 

\begin{equation}
\ket{\Phi} = \frac{1}{\sqrt{2}} \left( \ket{H_sH_i}_2 + e^{i\Delta\varphi} \ket{V_sV_i}_1 \right), 
\end{equation}

\noindent where $\Delta \varphi$ denotes the phase difference between the two emission processes. In order to observe one of the two maximally entangled Bell states, $\Phi^\pm$, (from now on we assume $\Phi^-$ for convenience) in type-I phase matching, the phase difference $\Delta \varphi$ of all collected photon pairs must be constant in all spatial and spectral degrees of freedom.


To determine the total phase difference, we consider the phase accumulated by the individual photons (e.g., the signal photon) as they pass through the elements of the setup,

\begin{equation}
\Delta\varphi(\lambda, \vec{x}, \vec{\alpha}) = \sum_j \varphi^V (\lambda, \vec{x}, \vec{\alpha}) - \sum_j \varphi^H (\lambda, \vec{x}, \vec{\alpha}).
\label{eq:phi}
\end{equation}

\noindent Here $\lambda$ denotes the photon wavelength, $\vec{x}$ the position within the crystal where the downconversion occurred, $\vec{\alpha}$ the signal emission angle and $j$ an index for each traversed medium (BBOs, air gaps, HWP, and post compensation crystal). The superscripts $H$ and $V$ denote the final photon polarization and also include the phase of the pump photons before downconversion (see supplementary material). The wavelength dependence of $\Delta\varphi$ is sufficiently compensated with a birefringent element\cite{trojekPhD,trojek2008collinear,villar2018experimental}(yttrium orthovanadate, YVO$_4$ in Fig.~\ref{fig:schematic}(a)). This leaves only angular and position dependencies, $\Delta\varphi = \Delta\varphi (\vec{x}, \vec{\alpha})$, which are usually negated by either using thin crystals or single-mode fiber collection. 


Regarding the position dependencies of $\Delta\varphi$, the design used in the present work has an advantage over previous configurations, e.g. using crossed crystals. 
In general, the SPDC pair origin of each emission process is mapped onto the transverse emission profile by the walk-off. For example, pairs originating from the entry face of a crystal are emitted from the top of the emission profile (see figure \ref{fig:schematic}(b), 1/i). These pairs traverse more nonlinear material than those originating from near the exit face of the crystal. Only if the phase difference for each point on the combined final transverse emission profile (at the exit face of the second crystal) is constant, the desired entangled state can be produced.

In the present configuration, phase compensation is achieved for each photon pair irregardless of the downconversion location within the crystals. For example, photon pairs generated at the entry face of the first crystal are spatially indistinguishable from photon pairs born at the corresponding position in the second crystal (see Fig.~\ref{fig:schematic}(b)). This results in a constant phase difference across the emission profile. The parallel axes configuration therefore allows the collection of SPDC photons through a field stop without single-mode filtering or severely restricting the collection region. This is a particular feature of the parallel axes configuration. In the anti-parallel case, it is straightforward to determine that $\Delta\varphi$ is not constant over the emission profile. 

The angular dependency of $\Delta\varphi$ is linked to the path lengths experienced by SPDC photons emitted under different angles (see Fig.~\ref{fig:schematic}(c)). This path length effect is shown in Fig.~\ref{fig:schematic}(d) as a function of the polar emission angles (see supplementary material for further details). 
To collect the maximally entangled Bell state one can restrict the SPDC collection angle appropriately using a circular field stop. Figure~\ref{fig:schematic}(e) illustrates how the relative phase impacts the entangled state fidelity.





The implementation of the entangled photon source is shown in Fig.~\ref{fig:setup}. The output from a collimated, narrow-band 405~nm laser diode ($\Delta \nu \leq 160$~MHz) is used to generate SPDC in the \SI{6}{\milli\metre} BBO crystals (optical axis angle $\theta = 28.8^\circ$) set for type-I, collinear phase matching. The photons are non-degenerate at 785~nm and 837~nm. The downconverted photons are collimated using a lens pair (see supplementary material), split according to their wavelengths and detected using Geiger-mode avalanche photo diodes (GM-APD). The spectral bandwidth of the photons is 25 nm for signal and idler photons and no additional spectral filtering is used.

To restrict the opening angles of signal and idler photons, a field stop can be placed in the collimated beam. We use a circular, adjustable iris as a field stop (Iris 1 in Fig.~\ref{fig:setup}). This allows the fidelity and brightness of the source to be actively controlled by the iris diameter. Alternatively, the iris can be placed in the signal or idler arm and scanned to access individual emission angles (Iris 2 in Fig.~\ref{fig:setup}).

To characterize an unknown quantum state, full quantum state tomography is usually employed. However, with some reasonable assumptions regarding the generated state, the measurement can be greatly simplified. As the phasematching prevents the generation of $\ket{HV}$ and $\ket{VH}$ components, the generated state is impacted only by the imbalance between the polarization components and the mixing of pure states with different values of $\Delta\varphi$. The fidelity of the photon pairs can be estimated by partial quantum state tomography using a single polarizer after the temporal compensation crystal. This projects signal and idler into the same linear polarization basis and, when the polarizer is rotated, this leads to a unique signature for the maximally entangled Bell states $\Phi^+$ (zero contrast curve) and $\Phi^-$ (full contrast curve) as discussed elsewhere\cite{villar2018experimental}.

\begin{figure*}[tb]
\centering
\includegraphics[scale =1]{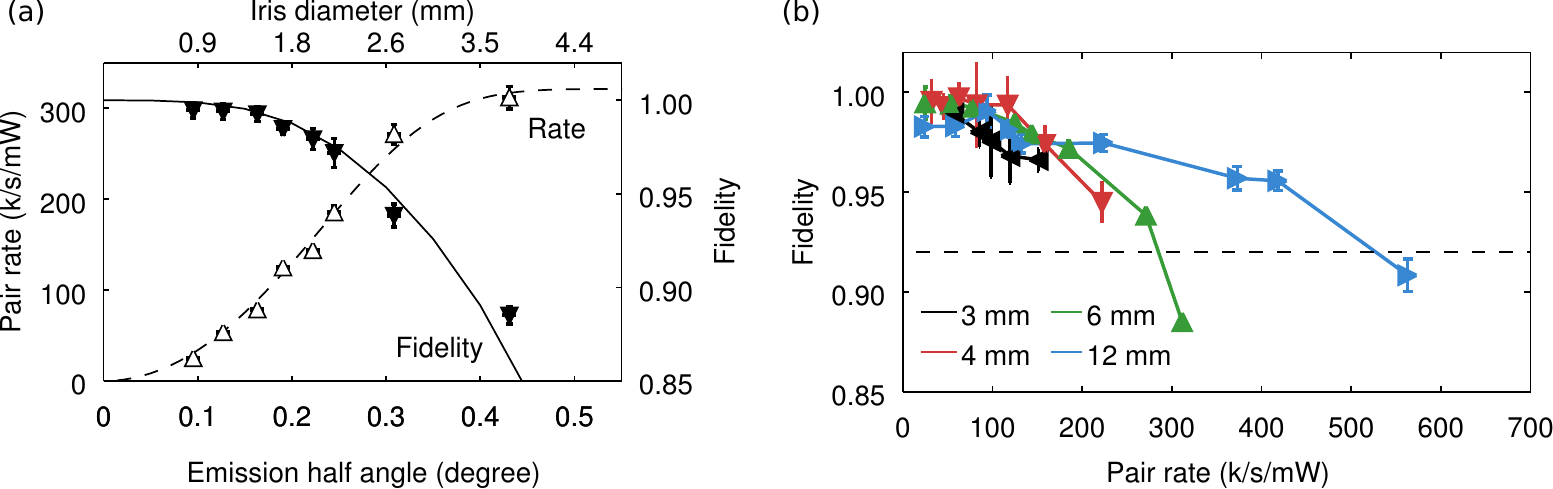}
\caption{(a) Observed pair rate (open triangles) and fidelity (filled triangles) as a function of emission angle for a source using 6~mm BBO crystals. The emission angle was calculated from the iris diameter. An error function was fitted to the pair rate data points. The solid line describes the calculated fidelity obtained from the phase map (Fig.~\ref{fig:schematic}(d), and is not fitted using a free parameter). (b) Correlation between fidelity and brightness for multiple crystal lengths. The brightness is controlled by the iris diameter. The dashed line indicates the QBER limit for the Ekert protocol of QBER$\leq15$\%\cite{gisin2007quantum}. (Color online)}
\label{fig:iris}
\end{figure*}

The angle dependent phase model can be validated by translating the iris in the signal arm while employing bucket detection in the idler arm. At each position of the iris, this setup enables the fidelity of the photon pairs emitted at different angles to be experimentally determined. Figure~\ref{fig:azimuth}(a) shows the characteristic signature of a state close to $\ket{\Phi^-}$ when the iris is in the center of the signal beam (iris diameter $d = 0.65\pm0.10$~mm). The contrast of the curve indicates a high entanglement quality and the fidelity towards $\ket{\Phi^-}$ extracted from the fit is $F=0.975 \pm 0.003$. We speculate that the mismatch from $\ket{\Phi^-}$ can be attributed to residual spatial and spectral phase components that impact the measured fidelity. Note that a fidelity of above 99~\% can be achieved as presented below.

The fidelity when the iris is translated across the signal beam in vertical direction is shown in Fig.~\ref{fig:azimuth}(b). In this case, the phase was set for maximum fidelity when the iris was at 0~mm and not adjusted for different data points. Ideally, the state transitions between $\ket{\Phi^-}$ and $\ket{\Phi^+}$ (see Fig.~\ref{fig:schematic}(e)). However, the fidelity is limited by the finite iris size and the consequent mixing of states with different values of $\Delta\varphi$. A model of the convolution of $\Delta\varphi$ with the finite iris size is constructed and shown in Fig.~\ref{fig:azimuth}(b) (solid black line). One standard deviation derived from the uncertainty of the iris diameter is shown as the gray shaded area. The experimental results and the model are in good agreement. In particular, the revival of the fidelity as the iris is translated in the signal arm confirms the relationship between phase and emission angle. This result holds for both, horizontal and vertical, polar angles (see supplementary material).

In the next step, we placed the iris in the collimated beam before splitting signal and idler photons and measured the pair rate as a function of the emission angle by gradually opening the aperture. The photon pair rate was measured at low pump power of $P\leq$\SI{100}{\micro\watt} to avoid saturating the passively quenched single photon detectors. The source brightness increased as expected for larger collection angles; with the fully open aperture, we observed a pair rate of $321\pm11$~\SI{}{k/s/mW} (see Fig.~\ref{fig:iris}(a)). As the angular far-field emission profile of the SPDC photons approximates a Gaussian function, the curve of the pair rate takes the shape of an error function. The observed pair-to-singles ratios were approximately constant over this range with $18.7\pm0.9$\% and $20.0\pm0.9$\% for signal and idler, respectively, and are mainly limited by the detection efficiency of the single photon detectors ($\leq 40-55$\%). In addition to the detection efficiency, the limitations for the heralding efficiency are losses due to reflection and absorption at optical surfaces.

For each iris diameter, we measured the fidelity. By opening the iris, we are effectively integrating over the variation in $\Delta\varphi$ as shown in Fig.~\ref{fig:schematic}(d). The relationship between fidelity and opening angle is shown in Fig.~\ref{fig:iris}(a). When the collection angle is below 0.1$^\circ$ the fidelity reached near-unity values of $F=0.995\substack{+0.005 \\ -0.007}$ and degrades for greater collection angles due to increased mixing. Such a degradation has been observed before for thin crystals\cite{kwiat1999ultrabright,altepeter2005phase}. When applied to thick crystals with collinear emission, our results show a high Bell state fidelity of $F\approx0.99$ at detected pair rates of more than \SI{100000}{pairs/s/mW} without pump shaping or spectral filtering. This is particularly interesting for applications outside of laboratory environments, where critically phase matched sources are thought to be advantageous due to their relative temperature stability. For comparison, a high-fidelity single-mode fiber coupled source based on a comparable crystal length and configuration yields pair rates of only up to \SI{65000}{pairs/s/mW} \cite{villar2018experimental}.


To further improve the brightness, one cannot simply increase the crystal length. This is because the angle dependent phase difference scales linearly with the interaction length (see supplementary material). The experimentally observed correlation between pair rate and fidelity is shown in Fig.~\ref{fig:iris}(b) for different crystal lengths.  As expected, higher pair rates can be collected from longer crystals under the same collection conditions, but there is always an upper limit for the pair rate at acceptable fidelity. 

One possible strategy to increase the pair rate while maintaining near-unity fidelity is to compensate the angle dependent phase difference. Due to the approximately circular symmetry of $\Delta\varphi$ (see Fig.~\ref{fig:schematic}(d)), this can be achieved conveniently by placing a spherical birefringent lens in the collimated beam. Such a lens could be used to simultaneously compensate any remaining spatial distinguishability caused by the different spatial origins of the $\ket{HH}$ and $\ket{VV}$~pairs.


In conclusion, we have shown that the beneficial spatial overlap in the parallel axes geometry enables the use of free-space detection which greatly enhances the rate of detected photon pairs for type-I, critically phase matched sources. This result was achieved for crystal lengths of up to 12~mm in contrast to earlier works which only focused on thin crystals\cite{altepeter2005phase,hegazy2017relative}. Similar to these previous works, the pair rate can be controlled by restricting the SPDC emission angle at the expense of entanglement quality. If high fidelity is needed, the pair rate may be reduced to ensure a near-unity fidelity, while for non-critical applications pair rates of more than \SI{0.4}{Mpairs/s/mW} are possible. Such detected pair rates are comparable to those in quasi phase-matched sources\cite{steinlechner2014efficient}.


In future work, the angle dependent phase may be compensated by an additional birefringent lens or a spatial light modulator to maintain the entanglement quality when the SPDC emission angle is not restricted. Moreover, an additional birefringent crystal may be added to the source to compensate the residual spatial mismatch between the $\ket{HH}_2$ and $\ket{VV}_1$ pairs. 

\section*{Supplementary Material}
The supplementary material contains further information about the phase calculations, the single polarizer measurement and the experimental setup.

\begin{acknowledgments}
This program is supported by the National Research Foundation (Award No. NRF-CRP12-2013-02), Prime Minister’s Office of Singapore, and the Ministry of Education, Singapore.
\end{acknowledgments}


\bibliography{bib}

\end{document}